\documentclass[conference]{IEEEtran}
\IEEEoverridecommandlockouts

\usepackage{cite}
\usepackage{amsmath,amssymb,amsfonts}
\usepackage{graphicx}

\usepackage{algorithm}
\usepackage{algorithmic}

\usepackage{fvextra}
\usepackage{listings}
\usepackage{xcolor}

\usepackage[T1]{fontenc}
\usepackage{lmodern}
\usepackage{textcomp}

\usepackage{xurl}
\usepackage[hidelinks]{hyperref}  
\hypersetup{breaklinks=true}
\usepackage{adjustbox}

\usepackage{balance}

\def\BibTeX{{\rm B\kern-.05em{\sc i\kern-.025em b}\kern-.08em
T\kern-.1667em\lower.7ex\hbox{E}\kern-.125emX}}

\begin{document}

\title{CTML: A Comprehensive Framework for Continuous Testing in Machine Learning Operations\\

}


\author{\IEEEauthorblockN{Bestoun S. Ahmed}
\IEEEauthorblockA{\textit{College of Engineering and Computing} \\
\textit{American University of Bahrain (AUBH)}\\
Riffa, Bahrain \\
bestoun.albeywanee@aubh.edu.bh}

}

\maketitle

\begin{abstract}
The increasing use of machine learning (ML) systems in production has highlighted some important weaknesses in traditional software testing methods. Unlike regular software, ML systems can fail in specific ways due to their reliance on data, including issues like data drift, concept drift, and performance decline. To address these issues, this paper introduces the Continuous Testing for Machine Learning (CTML) framework. This framework is designed specifically for MLOps environments and offers a thorough, automated solution. It establishes a standardized way to evaluate ML systems across four main areas: data quality, drift detection, model quality, and pipeline validation. The CTML framework consists of a modular setup that includes a Test Registry, Test Runner, Result Store, and Integration Layer. This design ensures it integrates smoothly with existing Continuous Integration/Continuous Delivery (CI/CD) pipelines and MLOps platforms. We put this framework to the test in a predictive maintenance case study, which showed how effective it is at identifying issues like data drift, performance degradation, and bottlenecks in the pipeline—problems that traditional testing methods might overlook. The findings indicate that CTML successfully uncovers critical failures in the system across all testing dimensions, providing useful insights to help keep ML systems running smoothly in production. Overall, this work presents a new, integrated framework that connects academic research with industry best practices, pushing forward the standards in MLOps quality assurance.

\end{abstract}


\section{Introduction}

Machine learning (ML) has fundamentally changed how we tackle complex problems across various fields. As organizations increasingly depend on ML for important business decisions, it is becoming essential to address the operational challenges that come with deploying and maintaining these systems. This shift has led to the emergence of MLOps (Machine Learning Operations), which adapts DevOps principles to meet the specific needs of the machine learning lifecycle \cite{singla2023mlops}. The unique features of ML systems create complexities that traditional software testing methods often can't handle. While conventional software relies on explicitly programmed logic, ML models learn and behave based on data through statistical methods. This data-driven approach introduces a distinct set of potential failure modes, making it necessary to implement specialized testing strategies.

The deployment and maintenance of ML systems come with challenges that set them apart from traditional software. Notable among these challenges are data drift and concept drift. Data drift refers to changes in the statistical properties of input data over time, while concept drift indicates that the relationships between input features and target variables can change \cite{BAYRAM2022108632}. These issues can lead to significant declines in model performance without any updates to the model code \cite{sahiner2023data}. In contrast to traditional software bugs that typically appear immediately, failures due to drift often develop slowly and can remain unnoticed for long periods.

Additionally, the performance of ML models heavily depends on the quality of the data they are trained on. Challenges such as missing values, outliers, inconsistent formatting, and shifts in the distribution of training or inference data can seriously hinder accuracy. Unlike traditional software, which typically employs straightforward input validation, ML systems deal with high-dimensional and complex data, where issues related to quality can be subtle and hard to detect \cite{boyko2024data}.

Moreover, ML systems consist of complex pipelines that involve various stages: data ingestion, preprocessing, feature engineering, model training, evaluation, and deployment. Each of these stages introduces potential points of failure, and the interactions among different components can result in unexpected issues that are difficult to anticipate. Standard integration testing often falls short when it comes to validating these intricate data flows and transformations.

Even though MLOps has made great progress in automating model training and deployment, testing is still a relatively underdeveloped part of the ML lifecycle. Most current practices focus on basic performance metrics—like accuracy, precision, and recall—evaluated against static test sets. However, this strategy is limited; static test sets do not reflect the evolving nature of production data. Models that perform adequately on historical data may show considerable decline in dynamic environments. Furthermore, standard metrics often provide an incomplete picture of model behavior, neglecting important aspects such as fairness, robustness, and generalization. Current testing efforts also tend to focus predominantly on the models themselves, overlooking the broader system context, including data pipelines, feature stores, and serving infrastructure.


To address the existing shortcomings in MLOps environments, we introduce the CTML (Continuous Testing for Machine Learning) framework as a comprehensive solution for continuous testing. This framework brings several important contributions to the table. Firstly, it establishes a structured testing methodology that spans four critical dimensions: data quality, drift detection, model quality, and pipeline validation. By doing so, it ensures thorough coverage of the entire ML system lifecycle. Additionally, the architecture of the framework is modular and extensible, allowing it to integrate seamlessly with established MLOps tools and CI/CD pipelines. This design allows organizations to adopt the framework incrementally, making it more accessible and adaptable. Moreover, the CTML framework implements automated continuous testing mechanisms that operate throughout the entire lifecycle of ML. This includes phases from development to production deployment and ongoing monitoring, ensuring that testing is maintained at all stages. The framework's effectiveness is further demonstrated through an empirical validation conducted via a detailed case study, which illustrates its ability to identify real-world issues across all defined testing dimensions. Lastly, the framework synthesizes insights from recent academic research and industry best practices, resulting in a solution that is both theoretically sound and practically applicable.


\section{Related Work}\label{relatedWork}

Data quality has re-emerged as a central reliability concern in modern data science and ML pipelines, where quality issues may go unnoticed until they surface as downstream performance degradation or unstable behavior. Dong et al. \cite{dong2024dqawareness} synthesize the evolution of data-quality awareness from traditional data management to data science systems, highlighting that ML pipelines introduce new quality failure modes (e.g., feature leakage, representation inconsistencies, and training/serving skew) that require dedicated, ML-aware checks rather than generic validation alone. This aligns with CTML's emphasis on continuous, automated data tests at both training and inference time, including distributional checks, schema/constraint validation, and actionable diagnostics.

The concept of MLOps emerged from the recognition that traditional DevOps practices, while valuable, require adaptation to address the unique characteristics of ML systems. Early foundational work by Karamitsos et al.\cite{karamitsos2020applying} explored how continuous automation principles from DevOps could be applied to machine learning, establishing the groundwork for more structured MLOps pipelines. This work highlighted the challenges of versioning data and models alongside code, a problem that does not exist in traditional software development. Building on this foundation, Garg et al.\ provided a detailed examination of CI/CD implementation for automated ML model deployment \cite{garg2021continuous}, while Liang et al.\cite{liang2024automating} presented a comprehensive overview of automating model training and deployment in MLOps, drawing extensively from industry best practices. Singla \cite{singla2023mlops} conducted a thorough analysis of MLOps challenges and strategies, highlighting the critical role of automated testing and continuous integration . However, these works treat testing as a secondary concern rather than a central pillar of the MLOps process—a gap that CTML addresses by elevating continuous testing to equal importance with continuous deployment.

Recent research focuses on detecting and understanding distribution shifts after deployment. Hinder et al. \cite{hinder2023survey} provide a survey on monitoring evolving environments, emphasizing that drift monitoring now requires not only detecting drift but also localizing and explaining it. Complementarily, Hinder et al. \cite{hinder2024feature} examine feature-based analyses for drift detection, proposing methods for semantic interpretation at the feature level. Monitoring frameworks for deployed ML models highlight that drift can impact features, predictions, and operational KPIs, necessitating systematic monitoring architectures over ad-hoc scripts \cite{eck2022monitoring}. Sahiner et al. \cite{sahiner2023data} review data drift in medical ML systems, indicating its implications for model reliability and patient safety, while Mallick et al. \cite{mallick2022matchmaker} provide evidence of data drift's severe effects in large-scale production systems through their Matchmaker system. The automation of drift detection is increasingly emphasized, with Mannapur \cite{mannapur2025understanding} reporting reductions in model degradation by up to 83.5\% and Rahik \cite{rahik2024automated} framing it as a strategic enabler for trustworthy AI. Additionally, Patchipala \cite{patchipala2023tackling} discusses the challenges of real-time drift monitoring systems, while Boyko \cite{boyko2024data} proposes anti-data drift strategies that focus on proactive data processing during model development. CTML incorporates these insights by embedding drift tests into a continuous testing workflow (test registration, execution, storage, and CI/CD integration), making drift signals essential outcomes that can trigger alerts, rollbacks, or retraining processes.

Recent software engineering research emphasizes that testing ML software must extend beyond static metrics to include fairness, robustness, and other trustworthiness criteria. Chen et al. \cite{chen2024fairnesssurvey} provide a survey of fairness testing, organizing efforts by workflows and components (data, model, and surrounding software), while highlighting open issues such as adequacy criteria and cost-effective strategies. Owen et al. \cite{owen2025validating} present a framework for validating AI/ML-based systems across multiple dimensions: model accuracy, fairness, robustness, and generalization. Balendran et al. \cite{balendran2025robustness} conducted a scoping review of robustness concepts in healthcare ML models, revealing inconsistencies in the definition and application of robustness testing. Their findings highlight significant gaps in the ML community's understanding and implementation of robustness, particularly in high-stakes domains like healthcare. These insights motivate CTML's approach of treating fairness and robustness checks as recurring tests conducted regularly, rather than as one-off audits. Moreover, empirical studies demonstrate that organizations often struggle to align research methods with practical testing activities, especially regarding data testing and metrics \cite{song2022mltestingpractice}. CTML addresses these gaps by defining automated test categories and integrating them into an operational workflow. Existing frameworks, while advancing individual testing dimensions, tend to be fragmented and fail to provide cohesive continuous testing throughout the ML lifecycle. CTML seeks to unify data quality checks, drift detection, and model quality testing into a comprehensive continuous testing paradigm.


\section{The CTML Framework}\label{ctml_framework}

To address the challenges identified earlier, we created the CTML\footnote{Source code is available at https://github.com/bestoun/ctml-framework} framework as a modular system for continuous testing in MLOps. It provides a structured approach to testing machine learning systems across four key dimensions and integrates seamlessly with existing MLOps tools.

\subsection{Framework Architecture}

\begin{figure}
\centering
\includegraphics[width=0.5\textwidth]{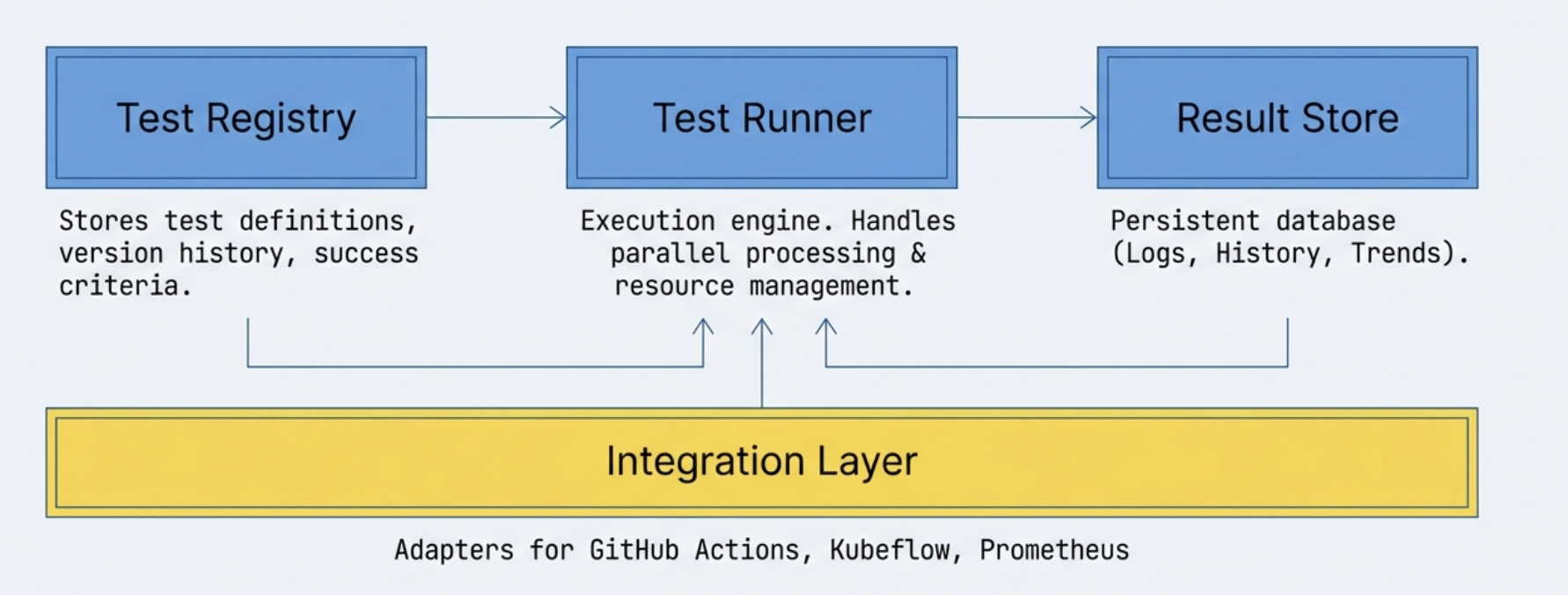}
\caption{CTML Framework Architecture}
\label{fig:ctml_architecture}
\end{figure}

The CTML framework is built around a core architecture comprising four primary components (Test Registry, Test Runner, Result Store, and Integration Layer) that coordinate testing activities across four testing dimensions as illustrated in Figure~\ref{fig:ctml_architecture}. This modular design enables flexibility and extensibility while maintaining a clean separation of concerns.

The Test Registry serves as a centralized repository for storing and managing test definitions. It provides a standardized format for defining tests across all testing dimensions, supporting versioning, metadata management, and test organization. The registry enables teams to share test definitions, track changes over time, and maintain a catalog of validated testing procedures. 

The Test Runner is the execution engine of the framework. It retrieves test definitions from the registry, executes them against the target ML system, and collects detailed results. The runner supports parallel execution of independent tests to minimize overall testing time, implements resource management to prevent system overload, and provides mechanisms for handling test failures gracefully. 

The Result Store is a persistent database for storing all test execution results, including detailed logs, performance metrics, and test artifacts. It provides a rich API for querying historical results, enabling trend analysis and the identification of recurring issues. The store supports multiple storage backends (relational databases, time-series databases, object storage) to accommodate different scale and performance requirements. By maintaining a complete history of test results, the Result Store enables teams to track system quality over time and identify degradation patterns before they impact production.

The Integration Layer provides the interfaces for connecting the CTML framework with external systems. It includes adapters for popular CI/CD tools (Jenkins, GitLab CI, GitHub Actions), MLOps platforms (Kubeflow, MLflow, SageMaker), and monitoring systems (Prometheus, Grafana, Datadog). This layer abstracts away the details of different tools, allowing the core framework to remain platform-agnostic. The integration layer also provides webhooks and REST APIs for custom integrations, ensuring that the framework can adapt to diverse organizational toolchains.

\subsection{Integration with the MLOps Workflow}

\begin{figure}
\centering
\includegraphics[width=0.5\textwidth]{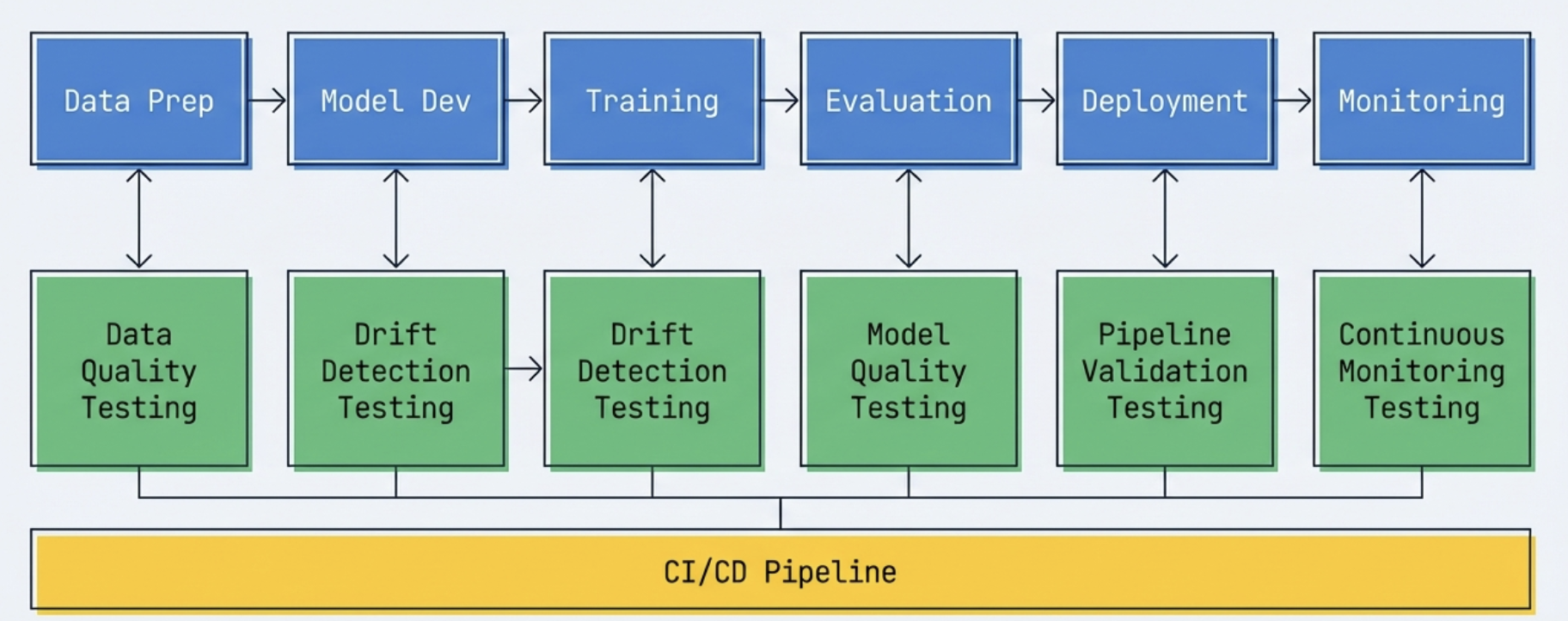}
\caption{Integration with MLOps Workflow }
\label{fig:mlops_integration}
\end{figure}

The CTML framework is designed to integrate at multiple stages of the MLOps workflow, ensuring that testing is a continuous process rather than an isolated activity. Figure \ref{fig:mlops_integration} illustrates how the framework integrates with the MLOps lifecycle, providing testing capabilities at development, CI/CD, continuous training, and production monitoring stages.

During the development phase, data scientists and ML engineers can use the framework to validate data quality and model performance locally before committing code. This early testing catches issues before they enter the CI/CD pipeline, reducing iteration time and preventing broken code from affecting other team members. The framework provides a command-line interface and Python API for local testing, making it easy to incorporate into existing development workflows.

When code is committed or a pull request is created, the CI/CD pipeline automatically triggers the CTML framework to execute a comprehensive test suite. This includes data quality checks, model performance validation, and integration tests. The framework provides rapid feedback on the impact of code changes, enabling teams to catch regressions before they reach production. Test results are reported directly in the CI/CD interface, with detailed logs and visualizations available through the Result Store.

Many modern ML systems employ continuous training, where models are automatically retrained on new data at regular intervals or when triggered by specific events. The CTML framework validates these newly trained models before they are promoted to production. This includes comparing the new model's performance against the current production model, checking for drift in the training data, and validating that the model meets all quality criteria. Only models that pass all tests are allowed to proceed to deployment.

After deployment, the framework continues to monitor the model's behavior in production. This includes real-time drift detection, performance monitoring, and anomaly detection. When issues are detected, the framework can automatically trigger alerts, initiate retraining workflows, or even roll back to a previous model version. 

\subsection{Testing Dimensions}

The CTML framework organizes testing activities into four key dimensions, each addressing a critical aspect of ML system quality. This multi-dimensional approach ensures comprehensive coverage of potential failure modes.

\subsubsection{\textbf{Data Quality Testing}} Data quality is foundational to ML system performance. The data quality testing dimension focuses on validating the integrity and suitability of data used for both training and inference. Completeness tests check for missing values in required fields, ensuring that the dataset contains sufficient information for model training or inference. These tests can be configured with thresholds (e.g., reject datasets with more than 5\% missing values) and can specify different requirements for different features based on their importance. Validity tests ensure that data values fall within expected ranges or belong to predefined sets of categorical values. For numerical features, this includes range checks (e.g., temperature must be between -50°C and 150°C), while for categorical features, this includes membership checks (e.g., country codes must be valid ISO codes). Validity tests help catch data corruption, sensor malfunctions, and integration errors. Consistency tests verify that data adheres to expected formats, patterns, and relationships, including format validation (e.g., dates must follow ISO 8601), pattern matching (e.g., email addresses must match a regex pattern), and cross-field validation (e.g., end date must be after start date). Consistency tests are particularly important for detecting data pipeline errors and schema evolution issues. Uniqueness tests identify duplicate records that could bias model training or inflate performance metrics, and can be configured to check for exact duplicates or near-duplicates based on similarity thresholds.

\subsubsection{\textbf{Drift Detection}} Drift detection is critical for maintaining model performance over time. The CTML framework implements both data drift and concept drift detection using statistical methods. Data drift occurs when the statistical distribution of input features changes over time. The framework implements multiple drift detection algorithms, including the Kolmogorov-Smirnov test (a non-parametric test that compares the cumulative distribution functions of two samples), the Population Stability Index (PSI) which measures how much a population has shifted over time, and Jensen-Shannon divergence (a symmetric measure of the difference between two probability distributions). These tests are applied to each feature independently, and the framework reports which features exhibit significant drift. By monitoring drift continuously, teams can detect when models need retraining before performance degrades noticeably. Concept drift occurs when the relationship between input features and the target variable changes. This is more challenging to detect than data drift because it requires access to ground truth labels. The framework monitors model performance metrics over time and uses statistical process control techniques to detect significant deviations from baseline performance.

\subsubsection{\textbf{Model Quality Testing}} Model quality testing evaluates the performance and behavior of the ML model itself. Performance tests evaluate standard performance metrics (accuracy, precision, recall, F1-score, AUC-ROC) against predefined thresholds. The framework supports different thresholds for different deployment environments (e.g., stricter requirements for production than for staging) and can compare new models against current production models to ensure that updates represent genuine improvements. Robustness tests assess the model's stability in the face of input perturbations through noise injection (adding Gaussian noise to input features to simulate sensor errors or measurement uncertainty), adversarial testing (applying small, targeted perturbations designed to cause misclassification), and out-of-distribution testing (evaluating model behavior on data that differs significantly from the training distribution). These tests help identify models that are overly sensitive to input variations, which could lead to unreliable predictions in production. Fairness tests evaluate whether the model exhibits bias across different demographic groups. The framework implements multiple fairness metrics, including demographic parity, equalized odds, and equal opportunity. These tests require careful configuration to specify protected attributes and acceptable fairness thresholds, but they are essential for ensuring that ML systems do not perpetuate or amplify existing societal biases.

\subsubsection{\textbf{Pipeline Validation}}
Pipeline validation tests the entire machine learning (ML) system rather than just the model alone. Integration tests check that all components of the pipeline work together, verifying data ingestion, feature engineering, model inference, and result delivery. These tests help identify issues like data format mismatches or API incompatibilities. Performance tests evaluate latency, throughput, and resource consumption, ensuring that the ML system meets production service level objectives (SLOs). They measure end-to-end latency and track CPU, memory, and GPU usage across the pipeline stages. Before deploying a new model version, deployment tests confirm that the process functions correctly, including model loading and prediction serving. These tests are vital for preventing deployment failures that could disrupt production.

\subsection{Test Execution Algorithm}

The core logic of the CTML framework is encapsulated in the test suite execution algorithm, which orchestrates the execution of multiple tests and aggregates their results. Algorithm~\ref{alg:execute_test_suite} presents the pseudocode for this process.

\begin{algorithm}
\caption{ExecuteTestSuite}
\scriptsize
\label{alg:execute_test_suite}
\begin{algorithmic}[1]
\STATE \textbf{Input:} $suite\_id$ (test suite identifier), $context$ (optional execution context)
\STATE \textbf{Output:} $suite\_result$ (aggregated test results)

\STATE \textit{// Initialize result object}
\STATE $suite\_result \gets$ new TestSuiteResult()
\STATE $suite\_result.suite\_id \gets suite\_id$
\STATE $suite\_result.start\_time \gets$ getCurrentTime()
\STATE $suite\_result.status \gets$ RUNNING

\STATE \textit{// Retrieve and validate test suite}
\STATE $test\_suite \gets$ TestRegistry.getSuite($suite\_id$)
\IF{$test\_suite =$ NULL}
    \STATE $suite\_result.status \gets$ ERROR
    \STATE \textbf{return} $suite\_result$
\ENDIF

\IF{$context \neq$ NULL \textbf{and} $\neg$ validateContext($context$)}
    \STATE $suite\_result.status \gets$ ERROR
    \STATE \textbf{return} $suite\_result$
\ENDIF

\STATE \textit{// Execute tests (sequential; "parallel" is descriptive only)}
\STATE $test\_results \gets [\ ]$
\FORALL{$test\_id \in test\_suite.tests$}
    \STATE $test\_def \gets$ TestRegistry.getTest($test\_id$)
    \STATE $result \gets$ TestRunner.executeTest($test\_def$, $context$)
    \STATE ResultStore.saveTestResult($result$)
    \STATE $test\_results$.append($result$)
\ENDFOR

\STATE \textit{// Aggregate results (use helper counters to avoid fragile set-notation)}
\STATE $passed \gets$ countPassed($test\_results$)
\STATE $failed \gets$ countFailed($test\_results$)
\STATE $errors \gets$ countErrors($test\_results$)

\STATE \textit{// Determine overall status}
\IF{$errors > 0$}
    \STATE $suite\_result.status \gets$ ERROR
\ELSE
    \IF{$failed > 0$}
        \STATE $suite\_result.status \gets$ FAILED
    \ELSE
        \STATE $suite\_result.status \gets$ PASSED
    \ENDIF
\ENDIF

\STATE \textit{// Finalize and store results}
\STATE $suite\_result.summary \gets \{passed, failed, errors\}$
\STATE $suite\_result.end\_time \gets$ getCurrentTime()
\STATE ResultStore.saveSuiteResult($suite\_result$)

\IF{$suite\_result.status =$ FAILED \textbf{or} $suite\_result.status =$ ERROR}
    \STATE NotificationService.sendAlert($suite\_result$)
\ENDIF

\STATE \textbf{return} $suite\_result$
\end{algorithmic}
\end{algorithm}

This algorithm showcases key principles of the CTML framework. It features strong error handling to ensure that failures in individual tests do not halt others. It also allows for parallel execution of independent tests, reducing total testing time. Additionally, it maintains detailed records of test executions for effective analysis and debugging, and it integrates with notification systems to inform teams of any issues.

\section{Evaluation Results and Discussion}\label{evaluation}

To validate the effectiveness of the CTML framework, we conducted a comprehensive case study using a predictive maintenance scenario. This section describes our experimental setup, presents results across all four testing dimensions, and analyzes the framework's ability to detect real-world issues.

\subsection{Case Study Setup}

We used a synthetic time-series dataset designed to simulate industrial equipment monitoring. The dataset contains sensor readings from multiple sources: temperature measurements (°C), pressure readings (PSI), vibration levels (mm/s), operational hours, and equipment failure labels (binary). The dataset consists of 10,000 samples collected over a simulated 12-month period, with approximately 15\% labeled as equipment failures, reflecting realistic class imbalance in predictive maintenance applications.

We trained a Random Forest Classifier with 100 trees to predict equipment failure probability. The model was trained on 70\% of the data, with 15\% for validation and 15\% for testing. The baseline model achieved 92\% accuracy, 0.88 precision, 0.85 recall, and 0.86 F1-score on the test set.

To comprehensively test the CTML framework, we created three datasets representing different real-world scenarios: (1) Baseline Dataset representing ideal conditions with clean data, (2) Drift Dataset with intentional distributional shifts (mean temperature shifted by +10°C and systematic bias added to sensor1) simulating seasonal variations or sensor calibration drift, and (3) Incomplete Dataset with artificially introduced missing values (approximately 10\% missingness) in temperature and vibration features, simulating sensor failures or data transmission errors.

\subsection{Data Quality Validation Results}

We executed the data quality test suite on the Incomplete Dataset to validate the framework's ability to detect data integrity issues. The framework successfully identified missing values in the temperature and vibration columns, as shown in Figure \ref{fig:data_quality_output}:

\begin{figure}
\centering
\includegraphics[width=\columnwidth]{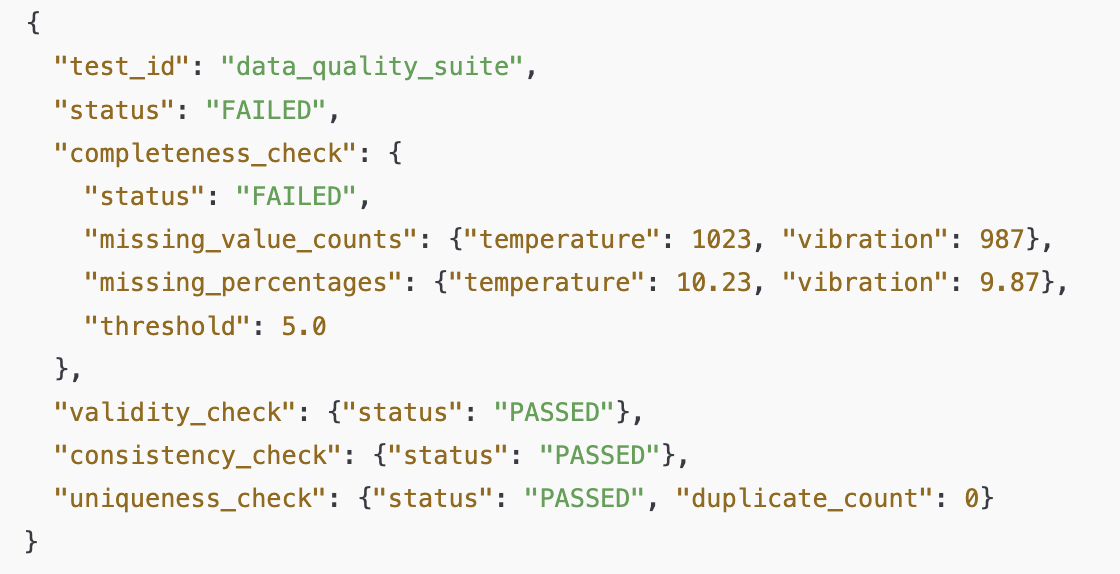}
\caption{Example output of automated data quality test suite}
\label{fig:data_quality_output}
\end{figure}

This result demonstrates that the data quality module effectively catches data integrity issues before they impact model training or inference. 

\subsection{Drift Detection Results}

\begin{figure}
\centering
\includegraphics[width=0.5\textwidth]{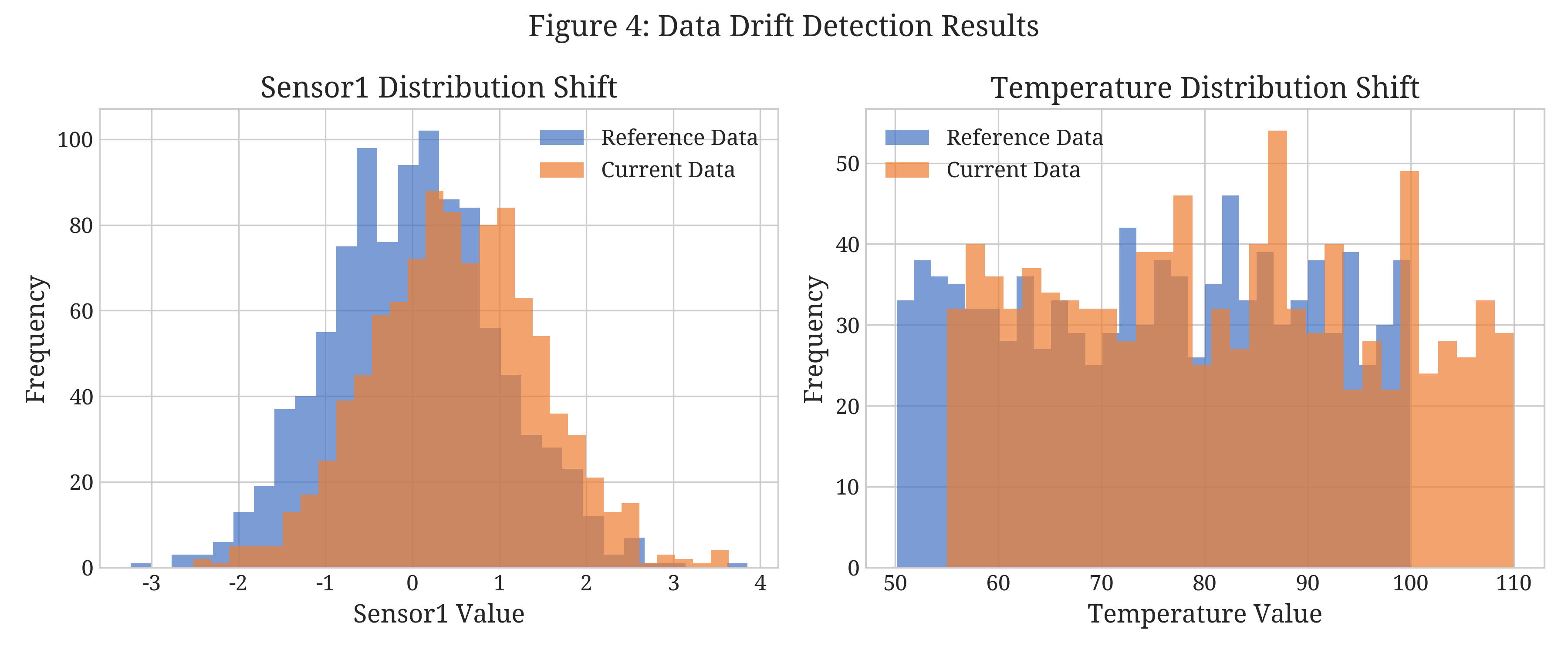}
\caption{Data Drift Detection Results Sample}
\label{fig:drift_detection}
\end{figure}

We used the drift detection module to compare the Baseline and Drift datasets, performing Kolmogorov-Smirnov tests on all numerical features to identify distributional shifts. Figure \ref{fig:drift_detection} shows histograms comparing the sensor1 and temperature distributions between the two datasets, with the CTML framework successfully detecting the noticeable shifts.

\begin{figure}
\centering
\includegraphics[width=\columnwidth]{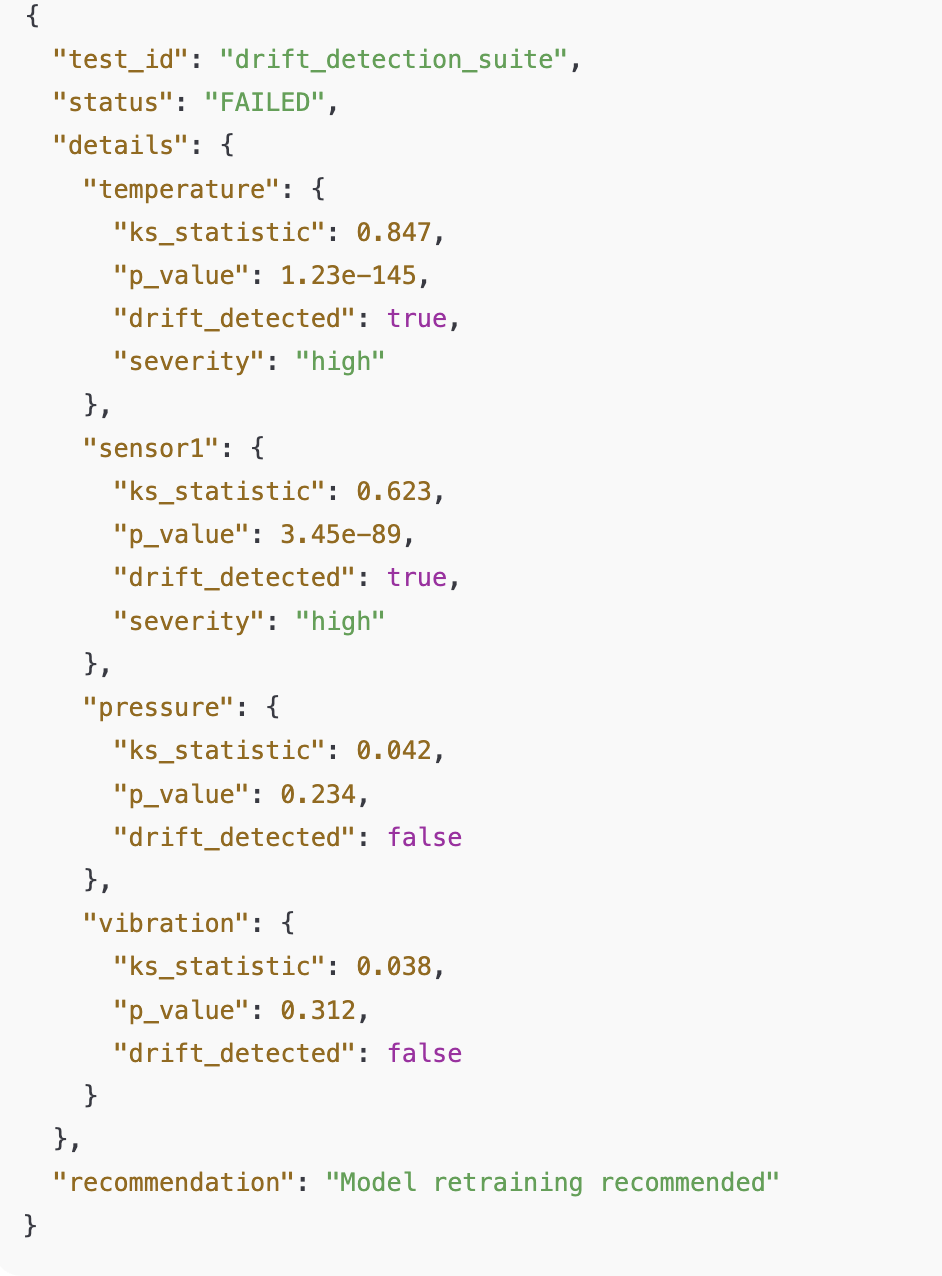}
\caption{Example output of drift detection test suite}
\label{fig:drift_detection_output1}
\end{figure}

The drift detection tests correctly identified significant drift in the temperature and sensor1 features as shown in Figure \ref{fig:drift_detection_output1}


These results validate the framework's ability to automatically detect data drift with detailed statistical information (KS statistic and p-value) providing quantitative evidence, while severity classification helps prioritize response actions.

\subsection{Model Quality Validation Results}

\begin{figure}
\centering
\includegraphics[width=0.3\textwidth]{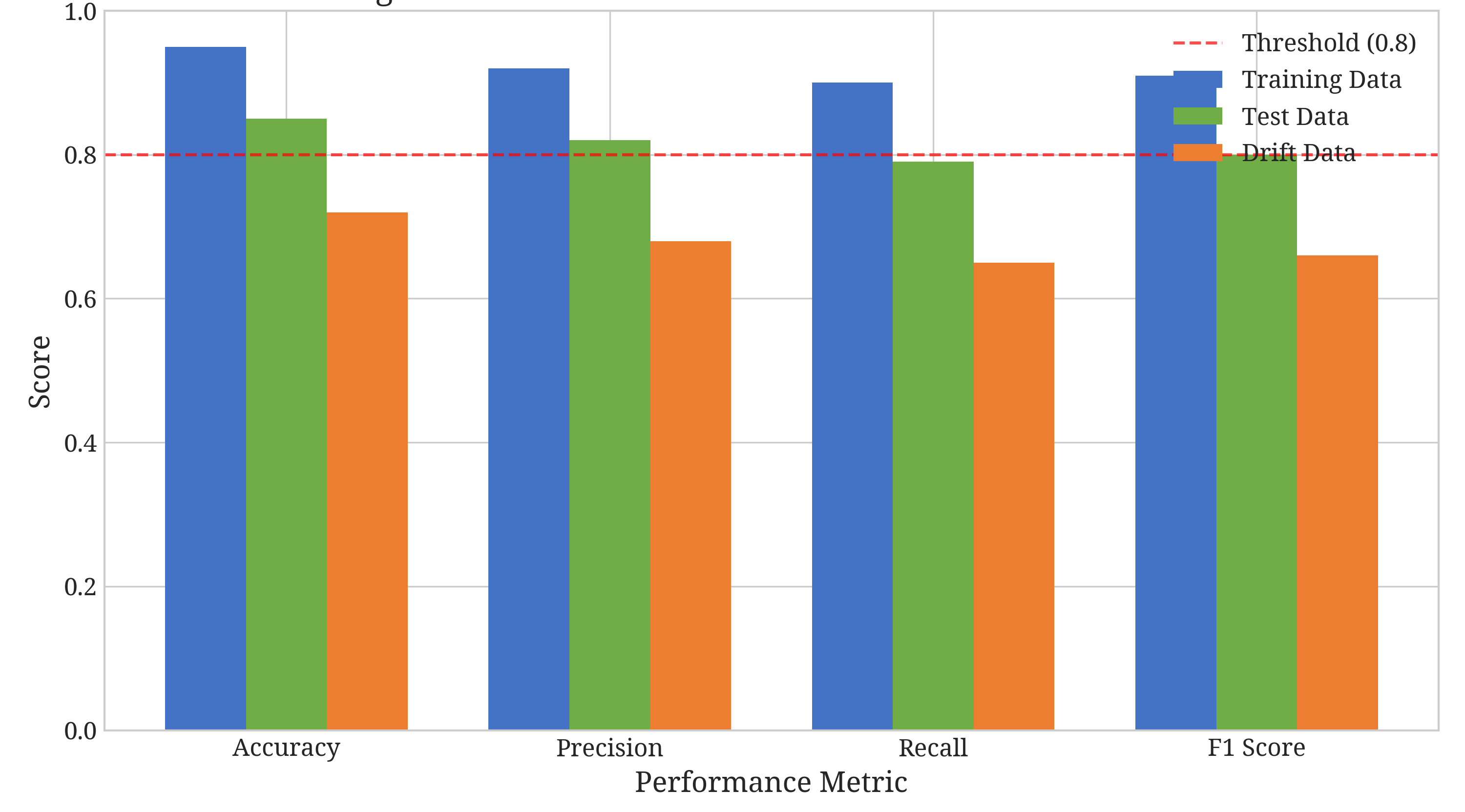}
\caption{Model Performance Comparison}
\label{fig:model_performance}
\end{figure}

We evaluated the baseline model's performance on both datasets to assess the impact of data drift on model quality, and conducted robustness tests by adding Gaussian noise to input features. Figure \ref{fig:model_performance} shows model performance comparison across baseline and drift datasets, showing accuracy, precision, recall, and F1-score. The significant performance degradation on the drift dataset demonstrates the impact of undetected data drift.

\begin{figure}
\centering
\includegraphics[width=0.3\textwidth]{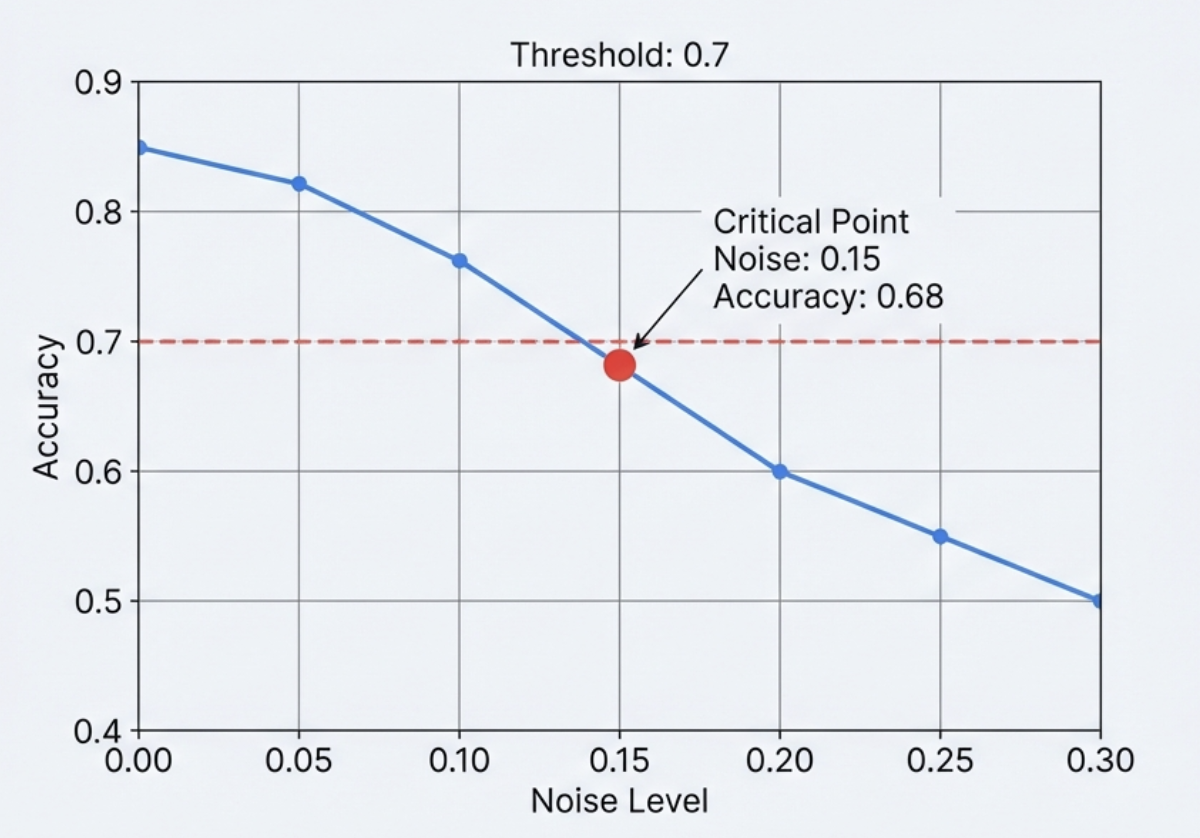}
\caption{Model Robustness Test Results}
\label{fig:robustness}
\end{figure}

We conducted robustness tests by systematically adding Gaussian noise with increasing standard deviations to input features. Figure \ref{fig:robustness} illustrates the degradation of model accuracy as increasing levels of Gaussian noise are added to input features. The red dashed line indicates the minimum acceptable accuracy threshold.

\begin{figure}
\centering
\includegraphics[width=\columnwidth]{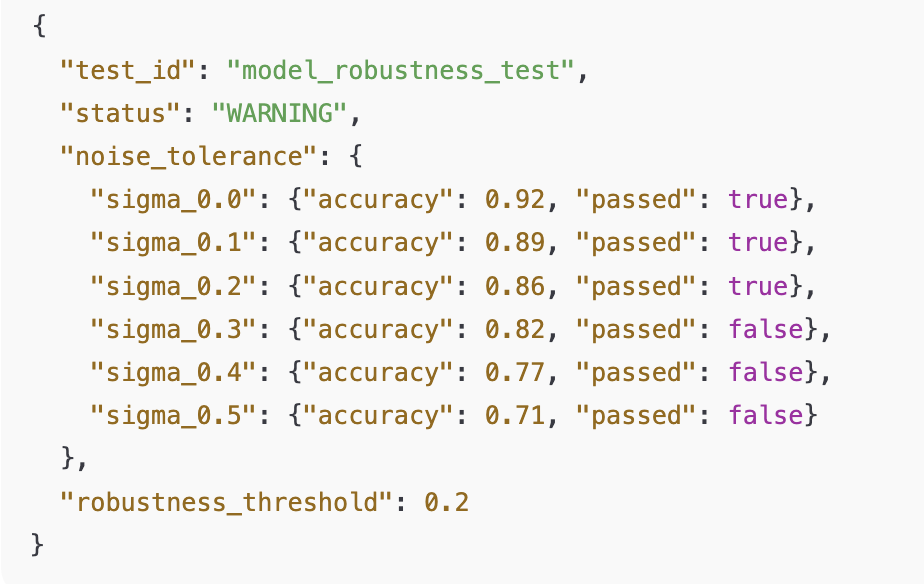}
\caption{Example output of model robustness evaluation}
\label{fig:model_robustness_output}
\end{figure}

The robustness tests revealed that accuracy falls below the acceptable threshold (0.85) when noise with a standard deviation of 0.3 or higher is added as shown in Figure \ref{fig:model_robustness_output}

These results demonstrate that the model quality module effectively identifies both performance degradation due to drift and robustness issues that might not be apparent from standard validation metrics alone.

\subsection{Pipeline Validation Results}

We executed pipeline validation tests to measure performance of the end-to-end inference pipeline, measuring execution time of each stage: data loading, preprocessing, feature engineering, model inference, and post-processing.

\begin{figure}
\centering
\includegraphics[width=0.35\textwidth]{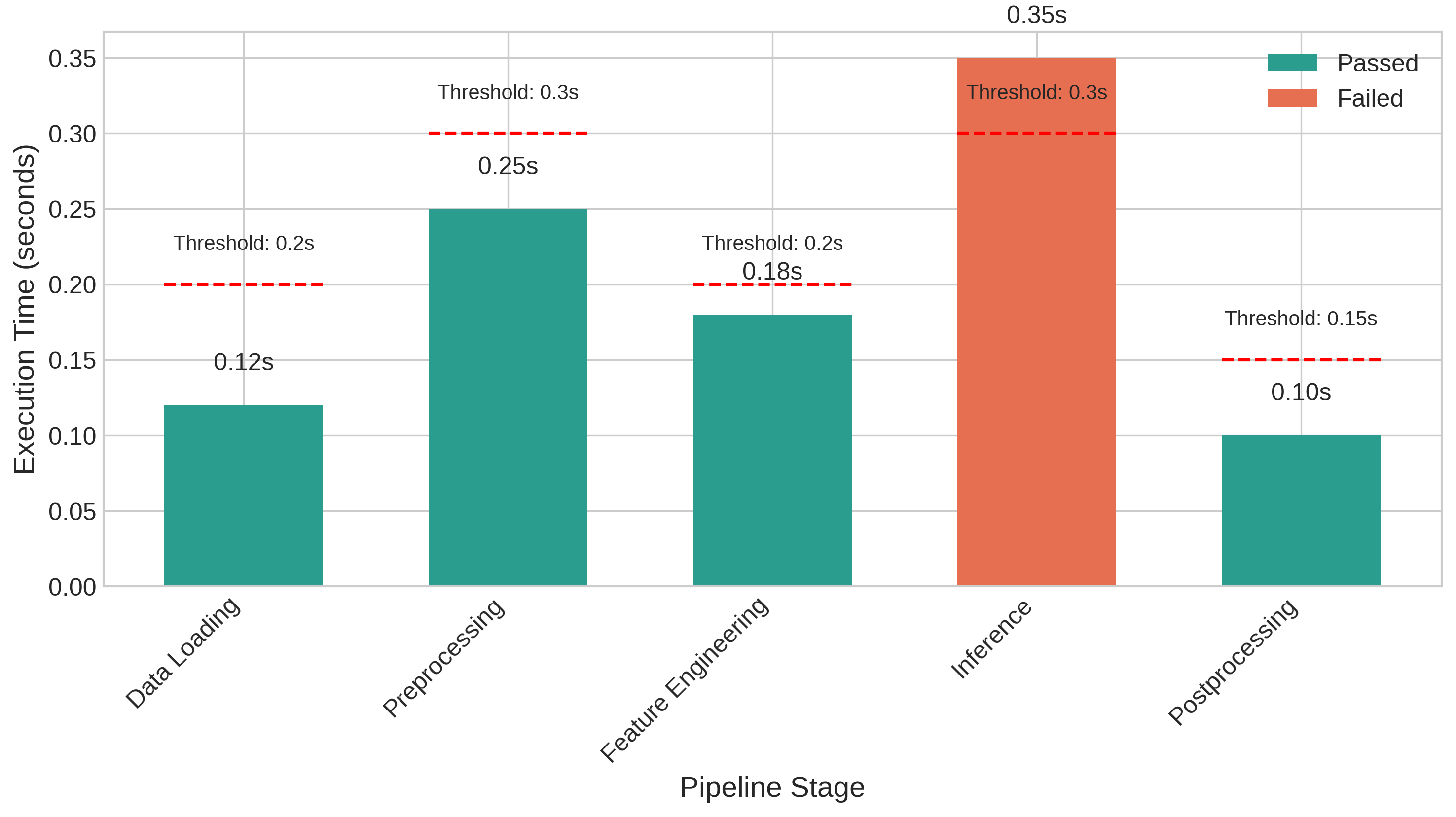}
\caption{Pipeline Performance Test Results showing execution time for each stage of the ML pipeline. The preprocessing stage exhibits significantly higher latency than other stages, indicating a performance bottleneck.}
\label{fig:pipeline_performance}
\end{figure}

This result highlights the importance of testing the entire pipeline, not just the model in isolation. The preprocessing stage accounts for nearly 75\% of total latency, suggesting optimization efforts should focus on this component.

\begin{figure}
\centering
\includegraphics[width=0.3\textwidth]{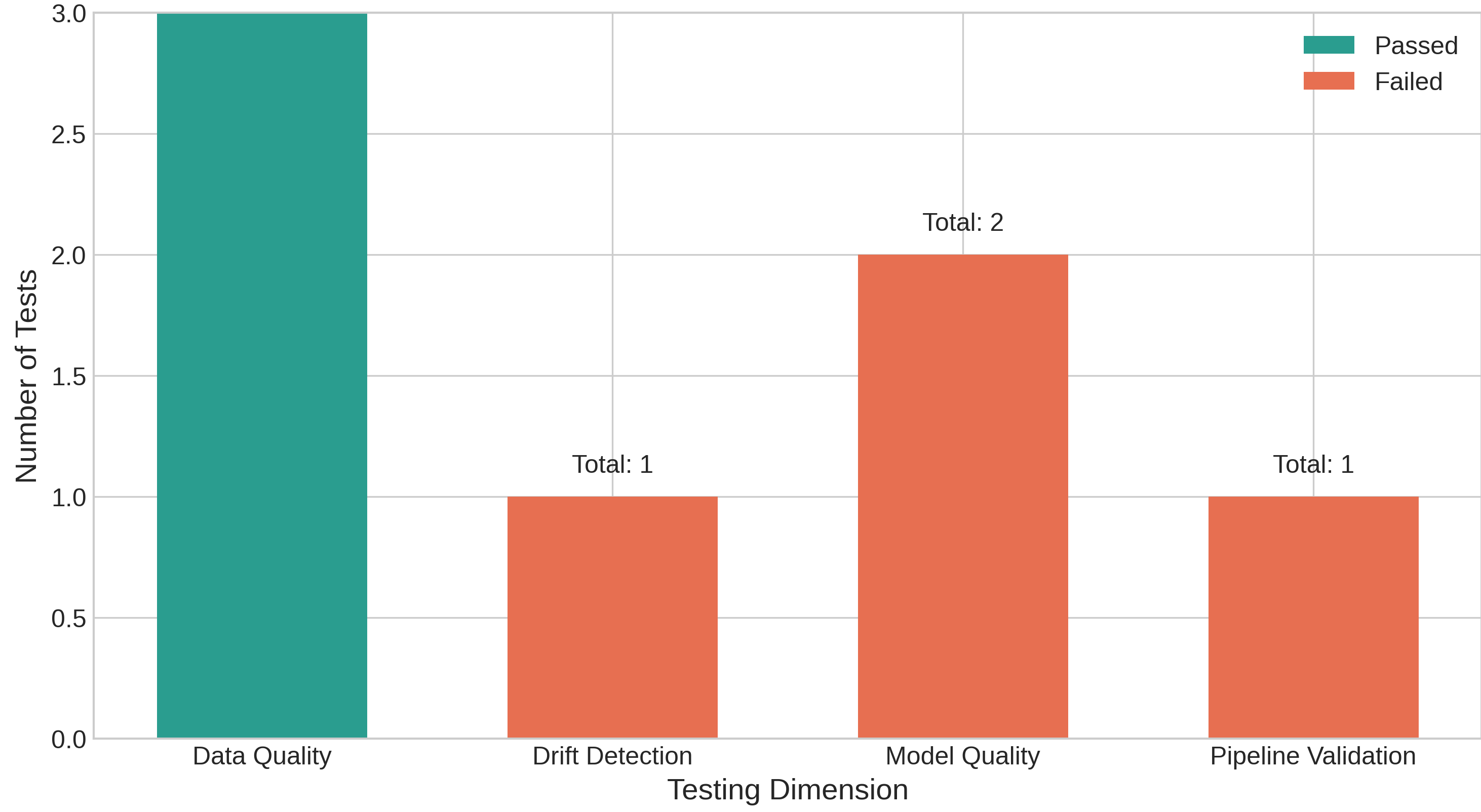}
\caption{Test Results Summary}
\label{fig:test_summary}
\end{figure}

The validation case study demonstrates that the CTML framework provides comprehensive coverage of ML system quality. By testing across multiple dimensions, the framework identified diverse issues that would have been missed by traditional testing approaches focused solely on model accuracy. Figure \ref{fig:test_summary} provides an overview of test results across all four testing dimensions, showing that the CTML framework successfully detected issues in drift detection, model quality, and pipeline validation while confirming data quality in the baseline dataset.

\subsection{Discussion}

The validation results strongly support the effectiveness of the CTML framework in identifying critical issues across all dimensions of machine learning (ML) system quality. Traditional ML testing often prioritizes performance metrics, such as accuracy or F1-score, based on static test sets, which provide an incomplete picture of system quality. Our case study illustrates that problems can arise in various areas: data quality issues can hinder predictions, while data drift can lead to silent performance degradation. Additionally, model robustness issues may only become apparent when exposed to noisy inputs, and pipeline bottlenecks can cause latency violations even when models perform well.

The detection of data drift highlights the necessity of continuous monitoring in production ML systems, as shifts in the environment can cause failures independent of code changes. Our findings show that data drift resulted in a significant accuracy drop from 92\% to 74\%, which could remain unnoticed without ongoing monitoring. The CTML framework addresses this issue with automated, continuous drift detection that enables proactive retraining. Additionally, robustness tests revealed that model performance significantly deteriorates when input noise exceeds specific thresholds, which is crucial for deployment decisions in noisy environments, such as those affected by electromagnetic interference. Traditional validation on clean data does not accurately represent model behavior under noisy or adversarial inputs, but the CTML framework’s robustness testing provides insights into model tolerance for input perturbations, facilitating more informed deployment choices.

Identifying a preprocessing bottleneck underscores the need to test entire ML systems rather than just individual models. In our case study, preprocessing accounted for 75\% of total inference latency, creating a bottleneck that model-centric testing could overlook, potentially leading to production latency violations. This aligns with industry reports that many ML production failures arise from pipeline issues instead of model problems.

The CTML framework provides key advantages over existing MLOps testing practices. Unlike tools like Evidently AI and Fairlearn, which focus on specific aspects like drift detection and fairness, CTML enables integrated testing across multiple dimensions. This reduces the reliance on disparate tools and offers a standardized method for defining, executing, and reporting tests, thus promoting best practices and consistent quality. With full automation from test execution to result notification, CTML minimizes manual effort and supports continuous testing throughout the machine learning lifecycle. Its Integration Layer allows seamless connections with existing MLOps tools and CI/CD pipelines, enabling easy adoption without workflow disruption.

However, some limitations exist. Our case study centered on predictive maintenance, and further validation across diverse domains (NLP, computer vision, recommendation systems) is necessary for greater generalizability. We used synthetic data for precise control, but real-world data is more complex. Our experiments involved relatively small datasets, and scalability to large systems handling millions of daily predictions is yet to be demonstrated. Currently, the framework relies on statistical tests for drift detection; incorporating advanced methods like learned drift detectors could enhance sensitivity. While basic fairness testing is included, more sophisticated metrics and debiasing techniques are needed for high-stakes applications. Lastly, the framework does not account for testing costs; resource-constrained environments may require prioritizing tests or adjusting their frequency.

\section{Conclusion}\label{conclusion}

This paper presents the CTML, an automated solution for continuous testing in MLOps. Traditional software testing is insufficient for ML systems due to unique failure modes like data drift, concept drift, and robustness issues. The CTML framework addresses these challenges through standardized testing in four key areas: data quality, drift detection, model quality, and pipeline validation. Our validation case study shows that the framework effectively identifies critical issues such as data quality problems, significant data drift, performance degradation, robustness vulnerabilities, and pipeline bottlenecks. By providing an integrated approach to ML testing, the framework bridges the gap between academic research and industry practice, enabling continuous testing throughout the ML lifecycle. As ML systems become vital to business operations and societal infrastructure, the CTML framework marks a significant step toward creating trustworthy systems where continuous testing is standard. Future work will include additional case studies with real-world data, advanced drift detection algorithms, expanded fairness testing, cost-aware testing strategies, libraries of pre-built test definitions, and integration with emerging MLOps platforms and cloud services.


\bibliographystyle{IEEEtran}  
\bibliography{ref}

\end{document}